%% file: main.tex
\documentclass[10pt,conference]{IEEEtran}
\IEEEoverridecommandlockouts
\usepackage{cite}
\usepackage{amsmath,amssymb,amsfonts}
\usepackage{algorithmic}
\usepackage{graphicx}
\usepackage{textcomp}
\usepackage{xcolor}

\usepackage{url}
\usepackage{multirow}
\usepackage{tikz}
\usepackage{caption}
\usepackage{subcaption}
\usepackage{booktabs}
\usepackage{svg}
\usepackage{dsfont}
\usepackage[shortlabels]{enumitem}
\usepackage[hidelinks]{hyperref}
\usepackage[colorinlistoftodos]{todonotes}
\usepackage[linesnumbered,ruled,vlined]{algorithm2e}

\def\BibTeX{{\rm B\kern-.05em{\sc i\kern-.025em b}\kern-.08em
    T\kern-.1667em\lower.7ex\hbox{E}\kern-.125emX}}

\begin{document}

\newcommand{\tobedone}[1]{\textcolor{blue}{#1\\}}
\newcommand\mycommfont[1]{\footnotesize\ttfamily\textcolor{blue}{#1}}
\newcommand{\shiv}[1]{\textcolor{red}{#1\\}}
\newcommand{\sarthak}[1]{\textcolor{green}{#1\\}}
\newcommand{\shagarw}[1]{\textcolor{orange}{#1\\}}
\newcommand{\ourmethod}{\texttt{ESRO}}
\newcommand{\etal}{\textit{et al.}}

\newcommand{\indep}{\perp \!\!\! \perp}

\makeatletter 
\newcommand{\linebreakand}{%
  \end{@IEEEauthorhalign}
  \hfill\mbox{}\par
  \mbox{}\hfill\begin{@IEEEauthorhalign}
}
\makeatother 

\title{\textit{ESRO}: Experience Assisted Service Reliability against Outages}

\author{\IEEEauthorblockN{Sarthak Chakraborty\IEEEauthorrefmark{2}\IEEEauthorrefmark{1}, Shubham Agarwal\IEEEauthorrefmark{3}, Shaddy Garg\IEEEauthorrefmark{4},
\\
\vspace*{0.1cm}
Abhimanyu Sethia\IEEEauthorrefmark{5}\IEEEauthorrefmark{1}, Udit Narayan Pandey\IEEEauthorrefmark{5}\IEEEauthorrefmark{1}, Videh Aggarwal\IEEEauthorrefmark{5}\IEEEauthorrefmark{1}, Shiv Saini\IEEEauthorrefmark{3}
}
\vspace*{0.1cm}
\IEEEauthorblockA{
\IEEEauthorrefmark{2}\textit{University of Illinois Urbana-Champaign, USA},
\IEEEauthorrefmark{3}\textit{Adobe Research, India}, \\ 
\vspace*{0.1cm}
\IEEEauthorrefmark{4}\textit{Adobe, India},
\IEEEauthorrefmark{5}\textit{Indian Institute of Technology Kanpur, India} 
\vspace*{0.1cm}
\\ sc134@illinois.edu, shagarw@adobe.com, shadgarg@adobe.com, 
abhimanyusethia12@gmail.com,\\
\vspace*{0.1cm}
 udit.pusp@gmail.com, videh1aggarwal@gmail.com, shsaini@adobe.com}
\thanks{\IEEEauthorrefmark{1}Work done while at Adobe Research, Bangalore}
}










\maketitle

\IEEEpubidadjcol

\begin{abstract}
Modern cloud services are prone to failures due to their complex architecture, making diagnosis a critical process. Site Reliability Engineers (SREs) spend hours leveraging multiple sources of data, including the alerts, error logs, and domain expertise through past experiences to locate the root cause(s). These experiences are documented as natural language text in outage reports for previous outages. However, utilizing the raw yet rich semi-structured information in the reports systematically is time-consuming. Structured information, on the other hand, such as alerts that are often used during fault diagnosis, is voluminous and requires expert knowledge to discern. Several strategies have been proposed to use each source of data separately for root cause analysis. In this work, we build a diagnostic service called \ourmethod{} that recommends root causes and remediation for failures by utilizing structured as well as semi-structured sources of data systematically. \ourmethod{} constructs a causal graph using alerts and a knowledge graph using outage reports, and merges them in a novel way to form a unified graph during training. A retrieval-based mechanism is then used to search the unified graph and rank the likely root causes and remediation techniques based on the alerts fired during an outage at inference time. Not only the individual alerts, but their respective importance in predicting an outage group is taken into account during recommendation. We evaluated our model on several cloud service outages of a large SaaS enterprise over the course of $\sim$2 years, and obtained an average improvement of 27\% in rouge scores after comparing the likely root causes against the ground truth over state-of-the-art baselines. We further establish the effectiveness of \ourmethod{} through qualitative analysis on multiple real outage examples.

\end{abstract}


\input{tex/01.Introduction}

\input{tex/02.RelatedWorks}


\input{tex/03.DataDescription.tex}

\input{tex/04.Overview}

\input{tex/05.GraphConstruction}


\input{tex/06.GraphInference}
\input{tex/08.Implementation}

\input{tex/09.Results}


\input{tex/10.Conclusion}

\bibliographystyle{IEEEtran}
\bibliography{main}

\end{document}

%% file: tex/01.Introduction.tex
\section{Introduction}

In recent years, software development and system design in organizations are moving away from traditional massive monoliths and towards a microservices-based design, resulting in faster rate of development and release~\cite{balalaie2016microservices, newman2021building}. These microservices are often deployed on the cloud and offered as Software-as-a-Service (SaaS) products to customers. However, this has raised concerns about maintaining the availability of these services since any production outage can negatively affect customers, resulting in significant financial losses for the enterprises~\cite{awsplan, azureplan}. For example, 37 minute of downtime on YouTube was estimated to cost Google US\$1.7 million in ad revenue alone~\cite{google_outage}, while one hour of downtime could cost Amazon US\$100 million on major shopping days~\cite{amazon_outage}. Despite ongoing reliability efforts over the years, cloud services continue to face unavoidable severe incidents~\cite{ghosh2022how, Amazon, Azure, Google} and outages~\cite{cloud_outages}. As a result, there has been a surge of research in the field of AI Ops (AI for IT operations)~\cite{cheng2023ai}.


In the traditional outage management workflow, Site Reliability Engineers (SREs) and On-Call Engineers (OCEs) manually investigate issues, leading to long investigation times and high resource wastage. This, in turn, increases both the mean time to detection (MTTD) and the mean time to remediation (MTTR), which are essential in maintaining service level agreements (SLAs)~\cite{Google_SRE}. The current outage management process consists of five steps:  (1) detecting outages through alerts, microservice traces, or performance metrics; (2) triaging the incident by communicating back and forth to assign the correct team for handling the issue; (3) identifying the root cause of the outage using multiple sources of data; (4) resolving the incident and finding a fix for the root cause; and (5) documenting the entire workflow as a natural language analysis report. This process is often inefficient and error-prone, requiring significant time and resources.


Diagnosing outages requires a significant amount of domain expertise, often gained from investigating past outages. However, manually searching through a large database of past outages is not feasible during the outage management process. As a result, additional resources are often required to communicate about any similar outages that have occurred in the past. Since most symptoms are not entirely unique, the expertise of SREs and the team responsible for a particular service is valuable throughout the entire process. Nevertheless, recent research studies, such as those presented in \cite{airalert, li2021fighting, NEURIPS2022_c9fcd02e, softner}, have explored data-driven techniques for predicting outages, performing root cause analysis, and triaging outages. These techniques can reduce the MTTD and MTTR while improving the overall On-Call Engineer (OCE) experience.

Several research studies, such as 
{We use `ancestral alert' instead of `root cause alert'\cite{li2021fighting}~\cite{li2021fighting, eWarn}, use alerts to detect and forecast outages, as well as suggest root cause alerts\footnote{We term alerts identified as `root cause alerts' in prior literature as `ancestral alerts' to avoid confusion with the outage reports terms.}. However, relying solely on alerts for root cause analysis can be inaccurate due to their volume and the presence of redundant and low severity alerts. Additionally, some alerts may not have triggered on the root cause service, but only on the affected services due to a snowball effect. Though some works use performance metrics collected through system monitoring to predict the root cause service, we argue that a larger set of data points will be needed since metrics capture normal and less critical system behaviour as well. Alerts are typically intended to identify problems that can have a significant impact on the system or its users, which makes it more suitable to be used to diagnose system failures. Some studies have also used past outage reports to predict likely root causes~\cite{salesforce_paper, ahmed2023recommending} by comparing symptom similarities. These works try to compute the similarity between the symptoms of the current outage and those described in previous outage reports. However, such a methodology will just map a recent incident to a past one, without considering any details regarding the pattern of the incident observed through alerts.

We conjecture that  utilizing both the alerts and the outage reports information in a systematic way leads to a more informed outage diagnosis process by assisting the OCEs. Though alerts are voluminous, they are structured and capture the real-time information about the degradation in the services, as the detailed symptoms may not be available until some time has elapsed since the fault occurred. Outage reports, on the other hand, contain rich information about past incidents from multiple services, but are written in semi-structured natural language text. Hence, combining this information with structured data can help engineers navigate and diagnose outages more effectively.


In this paper, we build an \underline{E}xperience assisted \underline{S}ervice \underline{R}eliability system against \underline{O}utages (\ourmethod{}) that retrieves similar outages based on a comparison of current alerts fired during an ongoing outage with symptoms of previous outages. It then recommends potential root causes and remediation techniques based on the retrieved outages. The novelty of our approach lies in integrating the structured information that is readily available in real-time during an incident along with the semi-structured historical outage reports to improve the experience of root cause analysis and performance diagnosis for the reliability engineers. The advantage that \ourmethod{} brings is the combination of outage-specific real-time information from the structured alerts data along with data-driven experience obtained from the past outages. We build a causal graph to represent the dependence relationships among the corresponding alerts, and a knowledge graph to represent the outage reports. The contribution of both sources of data is brought about by integrating the graphs in a way such that the alerts responsible for an outage are linked to the corresponding node represented in the knowledge graph. The linkage between the two graphs is accomplished through a temporal overlap of the outages with the alerts. To improve the prediction accuracy during inference, our approach also builds a predictor for an outage group (set of similar outages) using alerts, which ensures that only the alerts indicative of an outage are given more weightage in prediction. We have evaluated \ourmethod{} through real production outage data obtained from a large SaaS company, collected over a course of 2 years. We have observed at least 16\% improvement in accuracy in recommending potential root causes over state-of-the-art baselines. We further demonstrate the efficacy of our approach through real outage examples in the production scenario. 

The key contributions are summarized as follows:
\begin{itemize}
    \item We develop a system that contributes to the development of a more effective performance diagnosis methodology by leveraging structured and real-time alerts data along with outage reports which contain semi-structured natural language text.
    \item We build a causal graph to represent the alerts and a knowledge graph to represent the information present in the outage reports succinctly. We then merge the two graphs in a novel way to form a linkage between the alerts in the causal graph to the outage symptoms in the knowledge graph.
    \item Inference during an ongoing outage makes use of only the available alerts to rank past outages with similar set of alerts and symptoms.
    \item Experiments on real outage dataset shows the advantages of \ourmethod{} over the baselines. We observed 16\% improvement in predicting root causes and 38\% improvement in predicting mitigation steps. We also present qualitative review on few real outages.
\end{itemize}

%% file: tex/02.RelatedWorks.tex
\section{Related Work} \label{sec:related-work}

Root Cause Analysis has been studied in literature in the context of microservices and cloud services~\cite{wang2018cloudranger, ma2020automap, meng2020localizing, NEURIPS2022_c9fcd02e, eWarn, salesforce_paper, chakraborty2023causil,li2022causal}. Several works~\cite{wang2018cloudranger, meng2020localizing, causeinfer, NEURIPS2022_c9fcd02e, he2022graph} have utilized time series KPI metrics data obtained from \textit{Prometheus} to predict the root cause metric and service. These works usually build a causal graph among the performance metrics using some causal discovery algorithm~\cite{pcalgo}, which are traversed during inference time to locate root cause metrics. Qiu \etal \cite{mdpi_paper} on the other hand uses domain knowledge in the form of a knowledge graph to improve the causal graph learnt by the causal discovery algorithms and follows similar graph traversal algorithms to locate the root causes.

Works like AirAlert~\cite{airalert}, eWarn~\cite{eWarn} and Fog of War~\cite{li2021fighting} use alerts from multiple services for performance diagnosis. These works either extract suitable features from alerts or build a dependency graph structure for performance diagnosis of cloud services, and report the root cause alerts. However, in some cases, the root cause service may not trigger any alerts, making it difficult to correctly identify the faulty service solely through alert-based methods. Moreover, there is a possibility that the root cause service alert was triggered outside the designated time window used for creating alert-based features. In such cases, outage reports containing historical information provides more information about the root cause service, and the remediation technique that needs to be followed by comparing with similar past outages.

Works like \cite{salesforce_paper, softner, wang2021fast, ahmed2023recommending} mines information from the natural language text present in the outage reports for various performance diagnosis tasks. Saha \etal~\cite{salesforce_paper} builds a knowledge graph with the past semi-structured incident reports by extracting the symptom and root cause information from them using topic models~\cite{boudin2016pke} and language models~\cite{liuroberta}. It then runs inference on the graph to yield the most probable root cause. Ahmed \etal~\cite{ahmed2023recommending} used Large Language Models (LLMs) to understand the abilities of the outage report in predicting root causes and mitigation steps. Extensive experimentation with various language models suggests that outage reports are very useful in outage diagnosis. Another line of research uses the outage diagnosis data to learn correlations among them using deep-learning techniques to perform outage triaging~\cite{softner, wang2021fast, deepCT}. Liu \etal~\cite{liu2023incident} has attempted to correlate alerts with support tickets, but their design objectives differ from ours.

However, none of the above methods use any structured data available during a fault along with outage reports to recommend root cause and mitigation steps. Semi-structured information, such as outage analysis reports, is generated only after the fault has been mitigated or after a certain period of time following the occurrence of the outage. Thus, semi-structured data is unavailable during inference and hence we need to use structured alerts data which flows in real-time. The prior works do not address this issue.

%% file: tex/03.DataDescription.tex
\section{Data Description}


We give a brief overview on the different sources of data available to us for modelling \ourmethod{}. It uses structured data in the form of alerts obtained in real-time, along with the historical outage reports documented by the SREs after the mitigation of the outage.

\begin{enumerate}
\item \textbf{Alerts Data:} Alerts are fired by an alerting mechanism when the monitored metric values for a service component within a system exceed predetermined thresholds. These alerts can be of varying severity, and contain information such as the alert description, the condition that triggered the alert, severity level, the service affected, and the timestamp at which the alert was generated. For instance, a microservice may trigger an alert if the buffer queue size surpasses a certain value. The alerts data provide critical insights into the system's performance and any potential issues that may arise. They are available in real-time and forms the structured data.

\item \textbf{Outage Reports Data:}
Outage reports are created by the SREs after an outage has been resolved, through a comprehensive analysis and summary of the possible causes.  These documents capture the discussions during the outage and provide detailed insights into the resolution process. These reports provide a detailed description of the symptoms and its impact, as well as the root cause and remediation techniques employed. Additionally, the reports contain timestamps for when the outage was recorded and when it was resolved. Outage reports provide a duration during which the error in the system escalated, and hence the alerts generated during this period can be utilized to identify potential root causes. Moreover, by correlating outages with similar symptoms but for different system components, outage reports can suggest remediation techniques to mitigate similar occurrences in the future.
\end{enumerate}

\begin{figure}
    \centering
    \begin{subfigure}[b]{0.5\textwidth}
         \centering
         \includegraphics[width=0.9\textwidth]{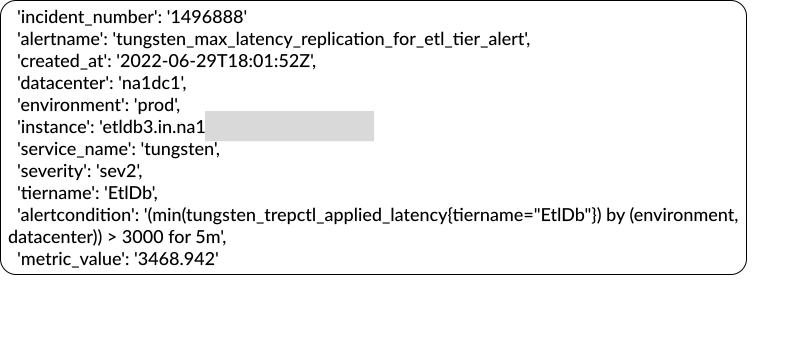}
         \caption{Structured Alert Data}
         \label{fig:alerts}
     \end{subfigure}
     \begin{subfigure}[b]{0.49\textwidth}
         \centering
         \includegraphics[width=0.9\textwidth]{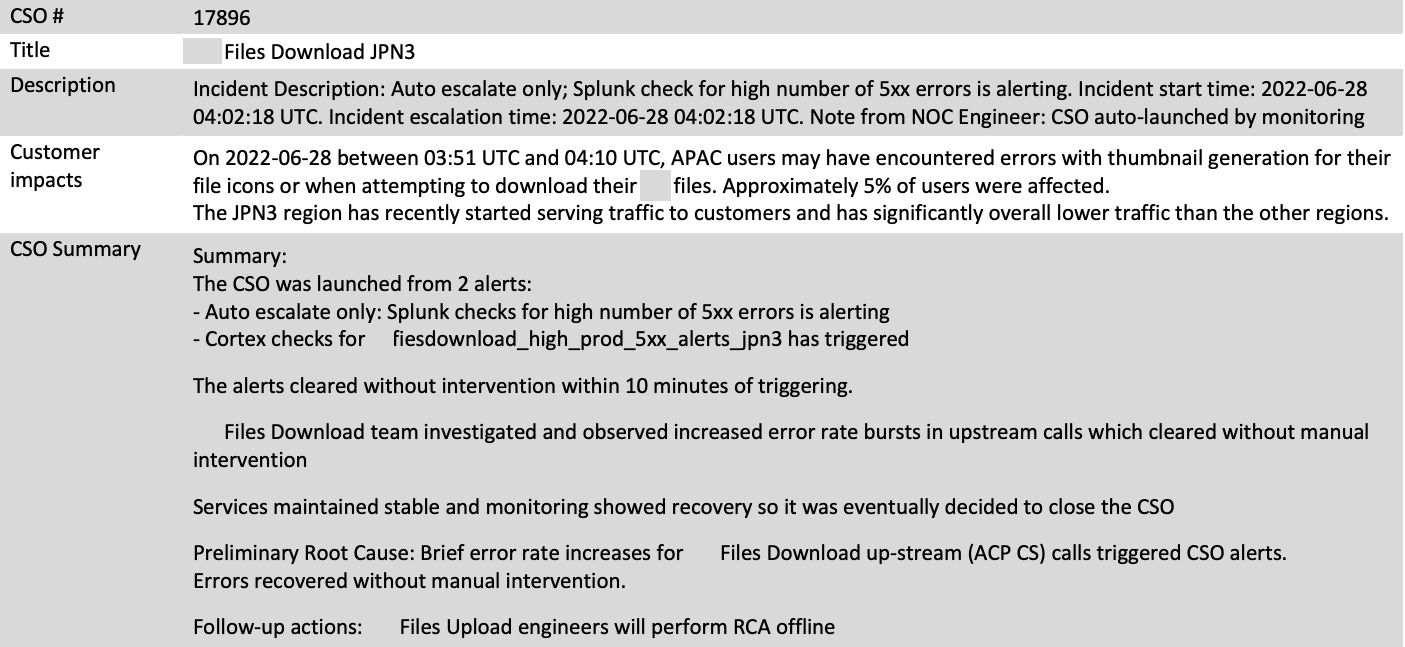}
         \caption{Semi-structured Past Outage Reports}
         \label{fig:CSO}
     \end{subfigure}
     \caption{\textit{Types of data available during an outage}}
    \label{fig:data}
    \vspace{-4mm}
\end{figure}

%% file: tex/04.Overview.tex
\section{Solution Overview} \label{sec:approach}


\begin{figure*}[t]
    \centering
    \includegraphics[width=\linewidth]{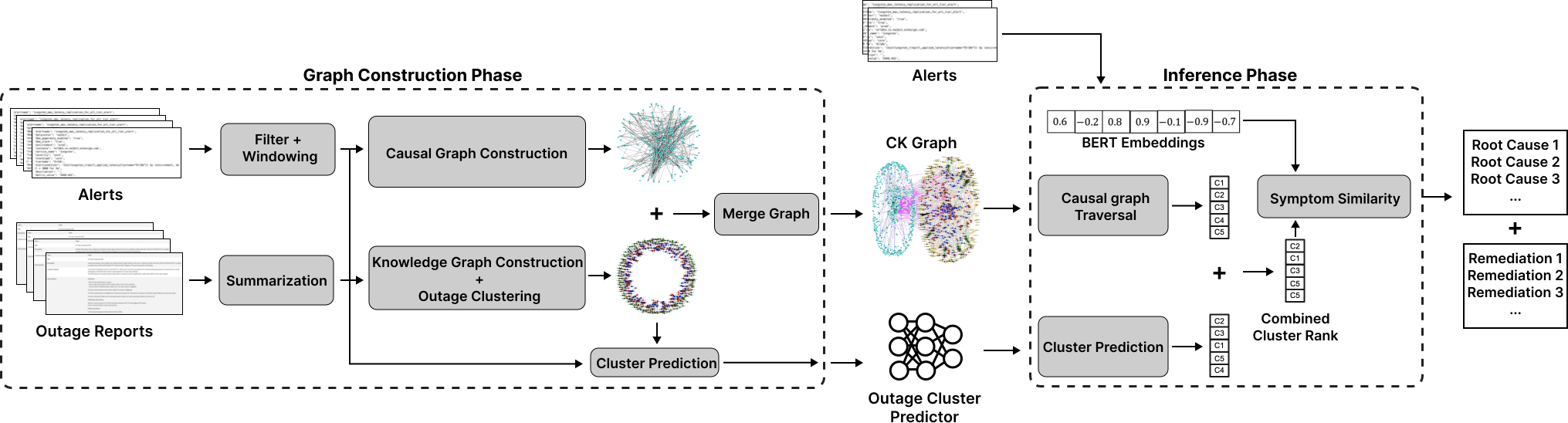}
    \caption{\textit{ESRO Pipeline consisting of two phases: (i) Graph Construction Phase - Previous alerts and outage reports are utilized to construct the merged CK Graph and train n outage cluster predictor, (ii) Inference Phase - Likely root causes and remediation steps are predicted from the real-time alerts for an outage}}
    \label{fig:our_approach}
    \vspace{-3mm}
\end{figure*}

Our approach involves building a system called \ourmethod{} that is capable of predicting potential and likely root causes and remediation steps during an outage. A schematic overview of \ourmethod{} is shown in Figure \ref{fig:our_approach}. 
Our proposed solution combines structured alerts data and semi-structured outage reports to accurately group similar outages that had occurred in the past and predict the likely root causes and remediation strategies from the set of similar past outages. The structured alerts data provides recent information on the system performance, while the semi-structured outage reports data offers detailed and comprehensive information on past outages. 


In order to capture the relevant information from structured alerts and outage reports data, we construct a causal graph (CG) and a knowledge graph (KG), respectively. The CG represents alerts as nodes and edges as causal relationships between alerts, while the KG summarizes the rich information present in the outage reports data. 
The training phase of \ourmethod{} merges the two graphs, creating a comprehensive understanding of system behaviour during an outage, which facilitates accurate prediction of root causes and remediation steps. This merging of the causal graph and the knowledge graph along with their use during inference time is a key novelty of \ourmethod{}, enhancing the effectiveness of our proposed solution.



Our approach involves: (i) construction of the individual and merged graphs, and (ii) leveraging the graph for inference at the time of an outage. Each of these steps can be further divided into sub-steps, which are described in detail. Experiments on real-world data demonstrate the effectiveness of our approach, which has the potential to significantly improve the reliability of complex distributed systems by enabling faster and more accurate resolution of outages.

%% file: tex/05.GraphConstruction.tex
\section{Graph Construction Phase} \label{sec:graph-construct}

In this section, we present the methodology for representing the information present in the alerts and the outage reports via graphs to facilitate root cause and remediation steps prediction. This involves multiple sub-tasks, including information extraction from incident reports and alerts, constructing individual graphs and finally merging the two graphs. Next, we delve into a detailed description of each sub-step. 


\subsection{Information Extraction} \label{sec:info-extract}
\begin{enumerate}
  \item \textbf{Alerts:} The set of alerts generated is grouped into the nearest time window of duration $t$ based on the timestamp at which they were fired. For our experiments, we have set $t$ to 15 minutes. This grouping results in a list of alerts fired for each time window. We then construct an indicator dataset, where each row represents the indicator function for each alert at a specific time window of $t$ minutes. The number of columns in the dataset equals the number of unique alerts fired, while the rows represent $t$ time duration windows. The structured dataset is considerable in size and potentially noisy, with certain columns exhibiting low variability and limited significance. Hence, we filter the number of time windows and the number of unique alerts to remove noise. Specifically, we remove the unique alert columns where alerts fired less than 10 times in the entire time period of over 1.5 years for which the data is collected, subject to the condition that no such alert fired during an outage. This ensures that only the relevant and frequent alerts are considered. We apply an additional filtering criterion to remove 95\% of time period rows where no alerts fired when there was an outage. This corresponds to the situations where the data does not show the relevant alerts.
  
    
    \item \textbf{Outage Reports:} The outage reports available are parsed to create a JSON, where symptoms, root cause, and remediation steps are the corresponding topics with their descriptions. Instead of utilizing the report's lengthy description of these attributes which contains various technical jargons, we aim to extract a summary that is more concise, comprehensive and free from domain-specific terminologies. To accomplish this, we use a pre-trained Bart-large summarization model~\cite{lewis2020bart} to extract a shorter summary for each outage's respective symptom, root cause, and remediation sections. We employ the abstractive summarization technique instead of the extractive summarization method for two reasons: (i) the original report contains reliability-specific jargon that makes it difficult for an extractive summary of an outage report to capture the details succinctly, and (ii) it can interpret information from multiple sources making it highly versatile in handling diverse and complex content (iii) it can condense lengthy and convoluted text which was more effective for our specific use case
\end{enumerate}

\subsection{Causal Graph (CG) Construction}
The indicator dataset of alerts occurrence data as obtained from Section \S\ref{sec:info-extract} forms the input to a standard PC algorithm~\cite{pcalgo} to obtain the causal graph between the alerts. Here, the alerts are the nodes in the graph and an edge $a\rightarrow b$ between two alerts $a$ and $b$ indicates that alert $b$ was caused due to alert $a$. 

PC algorithm is a constraint-based causal discovery algorithm that identifies the dependence relationships between pairs of alerts in the alerts occurrence data. It starts with a completely connected graph between the alerts, and iteratively computes the skeleton graph by removing relevant edges inferred through hypothesis testing using conditional independence (CI) tests. Here, we have used $\chi^2$ test since the data is discrete, that is, either the alert was triggered at a specified time or not. Specifically, PC algorithm runs CI tests of the form $p(y \indep x | S)$, where $x$ and $y$ are two alerts under consideration while $S$ is a set of alerts conditioned upon (also called separating set). The algorithm starts with an empty separating set ($S = \phi$) and increases the cardinality of $S$. Once the probability $p$ is greater than a confidence threshold $\alpha$, the PC algorithm removes the edge between $x$ and $y$. After constructing the skeleton graph, it then orients the direction of the edges using a set of rules~\cite{meek1995complete}. Hence, the result of employing the PC algorithm is a Completed Partially Directed Acyclic Graph (CPDAG), where the nodes correspond to distinct alerts. Within this graph, certain edges possess directed orientations, while others remain bidirectional. Bidirectional edges signify instances where the PC algorithm couldn't ascertain the causality direction between two nodes based on the available dataset.

Since the causal graph represents the causal dependence relationships among the alerts, an ancestral alert can be identified by traversing the graph. An ancestral alert is the one which was recorded first due to a fault in the root cause service, and it has in turn resulted in firing of other alerts in the impacted services. It should be noted that the service responsible for firing the ancestral alert might not be the root cause service. The causal graph thus can be used in real-time by traversing the alert nodes that were triggered during a fault.


\subsection{Knowledge Graph (KG) Construction} \label{sec:kg-construction}
The outage reports provide us with rich historical information of what were the symptoms, what were the root causes and how the symptoms were resolved. The entire information of the incident report can be represented through a knowledge graph with appropriate relations between the nodes. Hence, we construct a knowledge graph where the symptom, root cause and the remediation steps for each outage are represented as individual nodes. Furthermore, we add \texttt{has-root-cause} and \texttt{has-remediation} edge between each symptom and its corresponding root-cause node and remediation node extracted from the same outage report respectively.

However, such a graph would have separate connected components for each outage, implying similar outages will not be grouped together. As a result, the knowledge graph construction step also groups similar outages into a cluster. Based on the sentence embeddings of the corresponding abstractive summary of the symptom nodes, root cause nodes, and remediation nodes, we cluster them individually. We first tokenize each node description (symptom, root cause and remediation summary), and then use pre-trained contextualized BERT embeddings~\cite{kenton2019bert} (transformer based masked-language model) for each token/word in the summary to get the word embeddings. The advantage of using pre-trained BERT embeddings is that the embedding of each word is generated based on the context in which it has been used. We finally compute the node embedding by averaging the word embeddings for all the tokens/words present in the summarized node description. We refrain from using Sentence-BERT~\cite{reimers-2019-sentence-bert} since it requires structural and semantic flow in the sentence to obtain a high quality embedding. Such a semantic flow might not be present in the abstractive summary of the symptom, root cause or remediation.

For each individual node type (symptom, root cause and remediation), we performed separate Agglomerative Hierarchical Clustering~\cite{pedregosa2011scikit} on the node sentence description embeddings derived above. We then compute the optimal number of clusters $K$ based on the Silhouette score, such that the score with $K$ cluster is within 5\% of the maximum score possible. This is done in order to reduce the overall number of clusters. Finally, we combine the individual clusters formed for symptoms, root causes and remediation in such a way that two outages ($\mathcal{X}$, $\mathcal{Y}$) are grouped together in a cluster $\mathcal{K}$ if they were in the same cluster due to either their symptom ($\mathcal{K}_{symp}$), root-cause ($\mathcal{K}_{root-cause}$) or remediation ($\mathcal{K}_{rem}$).

\begin{equation}
    (\mathcal{X}, \mathcal{Y}) \in \mathcal{K} \Leftrightarrow (\mathcal{X}, \mathcal{Y}) \in \{\mathcal{K}_{symp} \vee \mathcal{K}_{root-cause} \vee \mathcal{K}_{rem}\}
\label{eq:cluster_KG}
\end{equation}

The intuition behind Equation \ref{eq:cluster_KG} is that an outage can be related to another if either their symptom are similar, or the root cause behind the incidents are similar or even if the remediation technique to mitigate the incident was similar. The knowledge graph hence represents the rich semi-structured information obtained from the outage reports.

\subsection{Merged Graph Construction} \label{sec:merged-graph}
The individual graphs (causal graph and knowledge graph) constructed above forms the basic components of the merged graph. Individually, they represent rich information but lacks usability. The process of merging combines the benefit of both the individual graphs and forms a comprehensive model to locate similar outages and recommend potential root causes and remediation techniques for an impending outage. We implement a novel mechanism to merge these two graphs, where we link the alerts in the causal graph to the symptoms in the knowledge graph. The idea behind this is that when a triggered alert in the system indicates some visible symptoms/anomaly in the system, the edge between the alert and the symptom reflects such a phenomenon. Since an outage report is not indicative of the exact set of alerts nomenclature that caused the outage, we use the timestamp of the outage and the alerts fired to link the causal and the merged graphs.

\begin{figure}[h]
    \centering
    \includegraphics[width=0.7\linewidth]{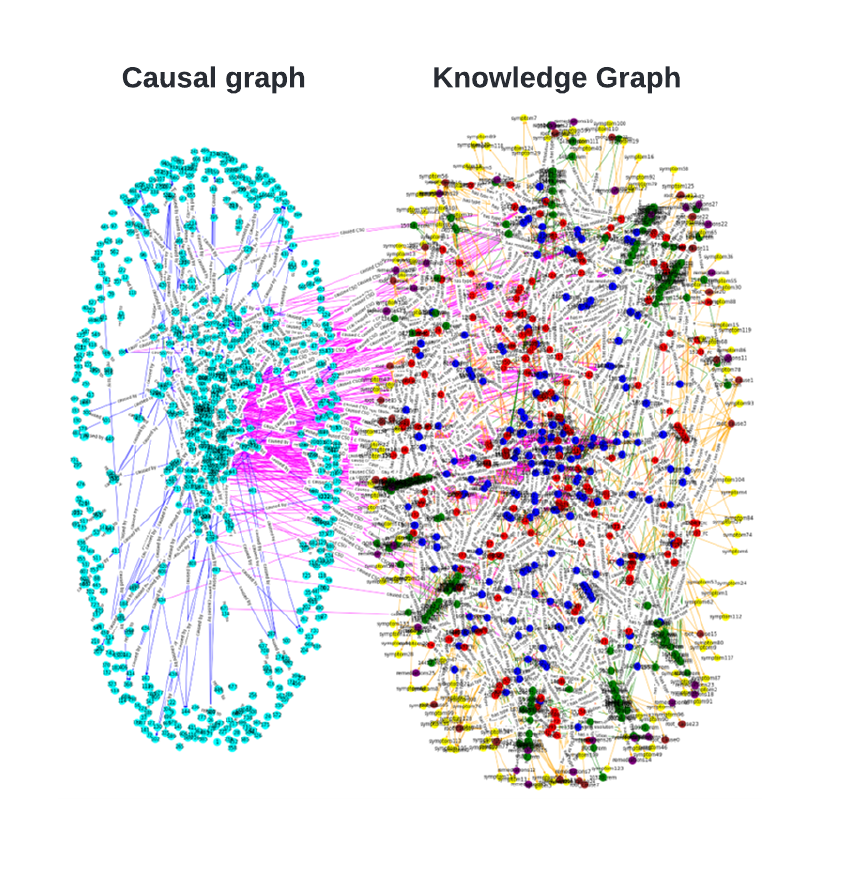}
    \caption{\textit{CK Graph after merging causal graph and knowledge graph. Pink edges connect the alerts in causal graph to symptom nodes in knowledge graph}}
    \label{fig:CG-KG}
    \vspace{-3mm}
\end{figure}

For each outage, we extract its start time and resolution time from the outage report. We then filter out all the alerts that were triggered at least once during the interval of the occurrence of outage or an hour\footnote{We select a duration of one hour for the outages as empirically we observed that most catastrophic issues happen within 1 hour after the root cause} before the start of the outage. These alerts are indicative of the outage and hence, if a combination of these alerts are triggered again, it is of high probability that a similar outage has occurred. Thus, for each alert in the list of filtered alerts, we add a \texttt{caused-outage} edge between the alert and the corresponding outage symptom node(s). Thus, we have the merged graph (Figure \ref{fig:CG-KG}), which we will henceforth term as \textit{CK graph}, which combines the structured as well as semi-structured data.

\subsection{Outage Cluster Predictor} \label{sec:cluster_predictor}
We link the causal graph with the knowledge graph based on the temporal overlap of the alerts fired during the occurrence of an outage. However, some alerts may have been fired that were not related to the outage that was occurring at the time. As a result, forecasting a single prior outage that is likely to occur when the alerts are triggered during inference time results in significant noise. Thus, we forecast a group of past outages by modeling only the most indicative alerts. However, the outage reports we have are not inherently clustered based on the set of alerts fired or the type of symptoms. Hence, we use the clusters that were defined in the knowledge graph (Section \S\ref{sec:kg-construction}) as ground truth to train a predictor to predict the cluster for an outage defined by a given set of alerts during inference. An additional advantage of using such a predictor is that it will inherently assign an outage prediction weight to each of the alert. Hence, not only the temporal overlap of the alerts with the outage is used, but also a weight is assigned to each alert, which will help the model to further rank the linked symptoms and their root causes.

Similar to Section \S\ref{sec:merged-graph}, we find the set of alerts triggered during the outage and an hour before the outage started, and create a dataset. However, the corresponding ground truth label that we want to predict is the outage cluster, which we computed in Section \S\ref{sec:kg-construction}. We then train a Random Forest classifier model on this dataset that predicts the cluster in which an outage belongs to, depending on the set of alerts fired. The maximum depth of the model is 25 while the number of estimators is 50. We utilize the inference of this model during the inference stage, which we shall describe below.

%% file: tex/06.GraphInference.tex
\section{Inference Phase} \label{sec:inference}

The constructed CK graph represents the temporal dependencies between the alerts and the past outages, with the similar outages clustered based on their symptoms, root causes and their remediation techniques. The CK graph is then  used by the inference pipeline to suggest possible root causes for a new outage based on the alerts triggered at the time, which is the only information available during inference. This realistic setting sets the approach apart from prior works that rely on symptom descriptions to infer potential root causes and remediation steps, which are not available until some time after the outage has started.

In this section, we describe multiple inferences methodologies, which have been compared in Section \S\ref{sec:results}. The various methodologies form our baselines while the Section \S\ref{sec:inference_clust} describes our proposed inference method. Below, we shall illustrate and/or exemplify the inference methodologies only for recommending the root cause, while analogous steps follow for the remediation techniques.

\subsection{Path-based Inference (Path)} \label{ref:path_inference}
\begin{figure}[h]
    \centering
    \includegraphics[width=0.75\linewidth]{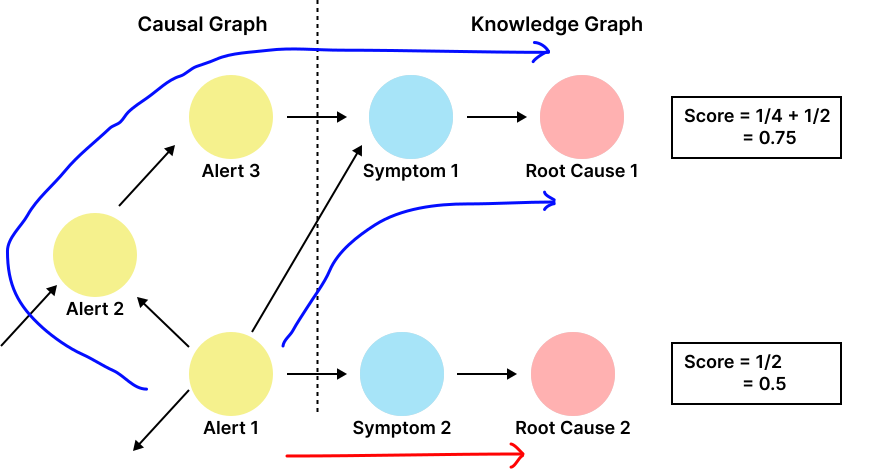}
    \caption{\textit{The figure shows a demonstration of the path based inference approach, where Alert 1 is fired during an outage. The inference method reaches two root causes Root Cause 1 (RC1) and Root Cause 2 (RC2) from Alert 1. There is only one 2-length path to RC2 from Alert 1, while there are two paths to RC1, a 2-length path and a 4-length path. Hence, the score for RC1 is 0.75 while the score for RC2 is 0.5.}}
    \label{fig:path-based}
    \vspace{-3mm}
\end{figure}

The causal graph component of the CK graph is traversed using a path based inference technique to find the collection of candidate ancestral alerts, with starting nodes being the alerts triggered during an outage. It leverages the structural and causal information in the alerts to predict likely root causes and remediation steps. A traversal of the causal graph will result in a collection of alert nodes that are directly linked to the symptom nodes in the knowledge graph component of the CK graph. We restrict our traversal to only those symptom nodes that are reachable in $k$ (or less) hops, where $k$ is a hyperparameter. The potential root causes are the corresponding root causes for the symptoms which had been reached though the traversal strategy. For each potential root cause $R$, a path-based score is defined.

\begin{equation}
    Score_{path}(R) = \sum_i \frac{1}{d(path_i \rightarrow symptom(R))}
\end{equation}

where, $d(path)=$ length of the path from the starting alert node to the root cause node. The inverse of $d(path)$ is summed over all the paths leading to the same root cause node. Figure \ref{fig:path-based} depicts demonstration of the same. A higher root-cause score represents a higher confidence in the prediction. This is based on two hypotheses: (i) the more the number of paths from the triggered alerts to a certain root cause, the higher is its likelihood of being the faulty service, and (ii) a shorter path represents a more direct correlation between the alert and the root-cause's historical co-occurrence and hence, should be given higher weight. This inference method utilizes only the information present in the structured data, that is, the alerts fired and their causal relationships to predict likely root causes.

\subsection{Similarity-Based Inference (Sim)} \label{sec:inference_similarity}
The similarity based inference method utilizes language processing techniques on the symptoms of the previous outages to predict likely root causes and remediation steps. It implicitly uses the information present in the semi-structured data, that is, the graph generated from the outage reports. 

We extract the title description of each triggered alert and use BERT to identify contextual embeddings for each word/token. The embeddings for all the words are averaged to compute the alert embedding. Finally, the average of the alert embeddings for all alerts triggered during the outage is used as an input query to the inference pipeline. We perform similar computations to obtain the symptom embeddings for each outage in the CK graph. Finally, we compare the cosine similarity of the input query embedding to each symptom embedding, and assign a score to each root causes with the similarity score for its corresponding symptoms. 

\begin{equation}
    Score_{Sim}(R) = cosine(emb_{alerts}, emb_{symptom(R)})
\end{equation}


Here $R\rightarrow symptom$ indicates the corresponding symptom of a root cause $R$ in the CK graph. With a higher score, a more similar symptom (from among the past outages) will be identified and hence, have a more similar root cause.

\subsection{Cluster based Inference (Clust)} \label{sec:inference_clust}
Our primary hypothesis is that combining the structured and current source of data (alert time series i.e. causal graph) with semi-structured and historical source of data (outage reports i.e. knowledge graph) for inference will result in better predictions than using each separately. In \textit{Clust}, \ourmethod{} utilizes the entire CK graph to predict the potential root causes and remediation techniques. 

Similar to the path-based inference method, \textit{Clust} traverses the causal links between the alerts to identify the set of ancestral alerts that link to the symptom nodes in the knowledge graph. It identifies all the set of symptom nodes that may be reached in $k$ or lower hops from the triggered alert nodes. However, unlike \textit{Path}, it reports a ranked list of the clusters to which the outages corresponding to the symptom nodes belong. The weight given to each cluster is the inverse of the number of times the cluster was reached using the graph traversal from the triggered alerts.

In the above set of clusters, we looked at the historical temporal overlap between the alerts fired and the outages to narrow down a set of potential outages to which the current outage is similar. However, to account for the importance of alerts fired in predicting the outages, we employ the outage cluster predictor described in Section \S\ref{sec:cluster_predictor}. We create a test set with indicator variables for all potential alerts using the alerts that were triggered during inference time and run the outage cluster prediction model to predict the probability of the alerts being connected with the clusters. 

We combine the two ranked list of clusters by adding up the individual weights. Finally, from the top-$L$ combined rank of clusters, we find the most similar symptom using the NLP based similarity technique as described in Section \S\ref{sec:inference_similarity}, and report the corresponding root causes and remediation steps. Thus, \textit{Clust} utilizes the entire CK graph in predicting the potential root causes and their remediation steps. Algorithm \ref{alg:inference_clust} describes the inference method.

\begin{algorithm}
\DontPrintSemicolon
\KwIn{CK Graph $\mathcal{G}$, outage cluster predictor $\mathcal{M}$, Set of all alerts $\mathcal{U}$, Set of fired alerts $\mathcal{A}$, $k$, $L$}
\KwOut{Potential Root causes and Remediations}

$Cluster\_Rank_1$ $\gets \{\}$ \;
$Cluster\_Rank_2$ $\gets \{\}$ \;
$\mathcal{C} \gets$ set of all outage clusters in $\mathcal{G}$\;
\vspace{2mm}
\For{each alert $a \in \mathcal{A}$}{
    Traverse $\mathcal{G}$ until $k$ hops to locate symptom nodes $\mathcal{S}$\;
    $Cluster\_Rank_1[get\_cluster(\mathcal{S})]$ += 1\;
}
Normalize $Cluster\_Rank_1$\;
\vspace{2mm}
$X \gets \mathds{1}_{\mathcal{U}}(a)$ $\forall a \in \mathcal{A}$\;
$Cluster\_Rank_2 \gets \mathcal{M}(X)$\;
\vspace{2mm}
$Cluster\_Rank[i] \gets Cluster\_Rank_1[i] + Cluster\_Rank_2[i]$ $\forall i \in \mathcal{C}$\;
$\mathcal{C}_L \gets$ top-$L$ clusters in $Cluster\_Rank$\;
\vspace{2mm}
$\mathcal{E} \gets create\_embedding(\mathcal{A})$\;
Ranked Symptoms $\mathcal{S'} \gets Sim(\mathcal{E}, emb_s) \text{ } \forall s \in \mathcal{C}_L$\;
$\mathcal{RC} \gets$ root causes for $\mathcal{S'}$\;
$Rem \gets$ remediation steps for $\mathcal{S'}$\;
\vspace{2mm}
\Return{$\mathcal{RC}, Rem$}
\caption{Clust Inference Method}
\label{alg:inference_clust}
\end{algorithm}

%% file: tex/08.Implementation.tex
\section{Experimental Setup} \label{sec:implement}
In this section, we outline the experimental process and the setup we followed. We have implemented \ourmethod{}\footnote{Data is proprietary and cannot be shared. Code is available at \href{https://github.com/sarthak-chakraborty/ESRO}{https://github.com/sarthak-chakraborty/ESRO}.} in \textit{python} and used causal-learn\footnote{\href{https://github.com/py-why/causal-learn}{https://github.com/py-why/causal-learn}}~\cite{causal-learn}, an open source library to construct the causal graph between the alerts. The natural language models used to compute the summary of the outage reports and create sentence embeddings were obtained from the open source implementation provided by Hugging Face~\cite{huggingface}. We have run \ourmethod{} on a system having Intel Xeon 8124M 3.0GHz CPU with 72 cores.

\subsection{Data} \label{sec:data-cso}
The research was performed on a dataset obtained from a large SaaS company that operates on a large-scale cloud infrastructure with thousands of servers and multiple data centres globally. \ourmethod{} focused on a production service that comprised of over 40 microservices deployed via Kubernetes, serving millions of users daily. The dataset contained alerts as well as outage reports for a period from October 2020 to June 2022, a period of almost two years.
\begin{enumerate}
\item \textbf{Alerts Data:} The data set contained $\sim44,000$ alerts logged by 13 distinct monitors. During the entire time period, $\sim940$ unique alerts were fired. However, after filtering the alerts data according to Section \S\ref{sec:info-extract}, only $330$ unique alerts remained for consideration, indicating that many alerts occur only a few times with no predictive power for an outage in the same service.
\item \textbf{Outage Reports:} The production level service outage reports comprised a total of 182 reports, with each report containing information on the symptoms of the outage, its impact, the root cause, and the remediation strategies involved in mitigating the consequences of the outage. There is also information on the affected services, the duration of the outage, and related incidents. Over the period of two years, there were around 85 unique symptoms and 95 unique root causes\footnote{While the exact manifestation of symptoms and root causes varies, it's worth noting that several instances may point towards analogous issues.}. 
\end{enumerate}

\subsection{Evaluation Methodology} 
\label{sec:eval_method}
We opt for `Leave-One-Out' strategy to test our model, which has been used in prior works~\cite{salesforce_paper}. While constructing the CK graph, we neither consider the outage report for the incident that need to be tested nor the alerts fired during the incident's duration. We retrieve top-k root causes and remediation steps of historical incidents where the symptoms were similar. It needs to be noted that we do not suggest a new root cause or a remediation step, but retrieve most relevant root causes and remediation steps for past incidents. We evaluate how close the predicted root causes and remediation steps are to the ground truth root cause and remediation steps respectively. Results in  Section \S\ref{sec:results} shows the evaluation metric's maximum value over top-k (here, k=5) predictions for all the different inference methodologies. Similar to the evaluation of a retrieval based task, success is essentially counted if there is a meaningful hit in its top-k predictions. We present an averaged evaluation over randomly selected 50 outages.

\subsection{Evaluation Metrics} 
We report Rouge (Recall Oriented Understudy for Gisting Evaluation)~\cite{lin2004rouge} score to report the similarity between the predicted root causes and remediation steps to the ground truth root causes and remediation steps respectively. Both of these are represented as text sentence. Rouge is used to compare a candidate text to a set of reference texts. Specifically, we choose Rouge-L and Rouge-1 scores~\cite{lin2004rouge}. Rouge-L takes into account sentence-level structural similarity and identifies longest co-occurring in sequence n-grams based on Longest Common Subsequence (LCS)~\cite{hirschberg1977algorithms}, while Rouge-1 measures the number of matching `1-grams’ between the two texts. Since the inference algorithm outputs a text prediction of the root causes and the remediation steps (and not a topic) based on a retrieval task, it might not match with the ground truth root cause exactly and hence a hit@top-k metric might not be a true metric. Hence, a rouge score will compute the closeness of the predicted root causes and the remediation steps to the ground truth.

We compare the similarity of the summarized text of the predicted root causes and remediation steps against the summarized text in the ground truth report, where the summary was computed/extracted using the methodology described in Section \S\ref{sec:info-extract}. The main reason for comparing the extracted summary to the ground truth is the presence of production level jargons in the entire text, which may affect the accuracy of our model. The extracted summary will capture the crux of the root cause, thus allowing a more relevant comparison. 

To predict the performance quality of the Outage Cluster Predictor model, we compute the top-K precision of its predictions at multiple values of K, where, we consider a prediction to be correct if the actual cluster is within the top-K predictions of the model.

\subsection{Baselines} \label{sec:baseline}
To evaluate the efficacy of our model, we implemented two baselines, which are the state-of-the-art approaches of utilizing outage reports~\cite{salesforce_paper}. Since no public code was available, we implemented them to the best of our understanding. The baselines are described as follows:
\begin{enumerate}
    \item \textbf{Incident Search (IS):} The symptom from the outage report for the test outage is used to search over the repository of all the remaining symptoms, by comparing similarity of the pre-trained RoBERTa~\cite{liuroberta} embeddings and top-5 similar symptoms are returned and evaluated. For this, we have used FAISS\footnote{\href{https://faiss.ai/}{https://faiss.ai/}}~\cite{johnson2019billion} which is a library developed by Facebook for efficient similarity search.
    \item \textbf{GCN:} 
    Symptoms, root causes and the remediation steps in the incident reports are represented as individual nodes with initial feature vectors being the average GloVe embeddings~\cite{pennington2014glove} of each word in their respective sentences. Edges between the different nodes (intra-symptom, intra-root cause, intra-remediation nodes and between symptoms and root causes) are adjusted as recommended by Saha \etal~\cite{salesforce_paper}. We train a 2-layered Graph Convolution Network (GCN)~\cite{kipf2017semisupervised} model with 16 dimension hidden layer followed by a Dense layer. We apply a contrastive loss $\mathcal{L}$ using 10 randomly sampled false root causes and the true root cause.
    \begin{equation}
        \mathcal{L} = -log(\frac{e^{sim(x_{s},x_{r})}}{e^{sim(x_{s},x_{r})} + \sum_{j=1}^{10}e^{sim(x_{s},x_{r,j}-)}})
    \end{equation}
    where $x_r=$ GCN representation of root cause, $x_s=$ GCN representation of symptom, $x_{r,j^{-}}=$ GCN representation of false root cause.
    
    During inference, we compute the cosine similarity of the input symptom's GCN representation with all the symptom representations in the graph. Corresponding root causes and remediation steps for symptom nodes having maximum cosine similarity is considered as output.
\end{enumerate}

Both of the aforementioned baseline methods utilize the knowledge graph component to predict both root causes and remediation steps. In the case of Incident Search, RoBERTa embeddings are employed to represent symptoms, root causes, and remediation strategies. Meanwhile, GCN employs a graph convolution network to train embeddings for knowledge graph nodes, enabling the retrieval of similar symptom nodes for a given incident. These baselines are aligned with our study as they try to retrieve the most relevant root causes and remediation options.
Given that outage reports exclusively provide the information about the actual root cause and the associated remedial actions for an outage, approaches that rely only on the causal graph of alerts is unsuitable as a baseline. In these approaches~\cite{airalert, 273869}, only the root cause alert can be retrieved from the causal graph, but without information about the actual root cause and the associated remedial actions it is neither useful nor can it be evaluated with ground truth. Therefore, we opt not to utilize baselines~\cite{airalert, 273869} that rely solely on alerts to predict root causes.

\subsection{Design Choices and Hyperparameters}
\begin{itemize}
    \item The optimal number of clusters in Section \S\ref{sec:kg-construction} are as follows: (i) $\mathcal{K}_{symp} = 63$, (ii) $\mathcal{K}_{root-cause} = 100$, (iii), $\mathcal{K}_{rem} = 112$, and (iv) $\mathcal{K} = 53$.
    \item $k=9$ in Section \S\ref{ref:path_inference}, i.e., the path from the alert nodes to symptom nodes of outages is of 9 or fewer hops, and hence path to respective root cause node is 1 additional hop
    \item $L=3$, that is, we choose top-3 clusters based on the combined ranking in the cluster-based inference method.
\end{itemize}

%% file: tex/09.Results.tex
\section{Evaluation Results} \label{sec:results}

We evaluate our model with all the inference methods listed in Section \ref{sec:inference} along with the baselines listed above. The evaluation strategy will be presented in the following way:
\begin{enumerate}[A)]
    \item Evaluating design choices in \ourmethod{}
    \item Comparison of the chosen design for \ourmethod{} against the baselines
    \item Measuring the quality of the outage cluster predictor
    \item Examples using actual production outages
\end{enumerate}

\subsection{Design Choices Evaluation}
In this section, we shall evaluate two principle design choices. One is computing the optimal number of clusters for grouping the outages during graph construction phase, while the second choice is for the most suitable inference method (Section \S\ref{sec:inference}) of \ourmethod{}.

\subsubsection{Number of Clusters}
\begin{figure}[h]
    \centering
    \begin{subfigure}[b]{0.49\linewidth}
         \centering
         \includegraphics[width=\linewidth]{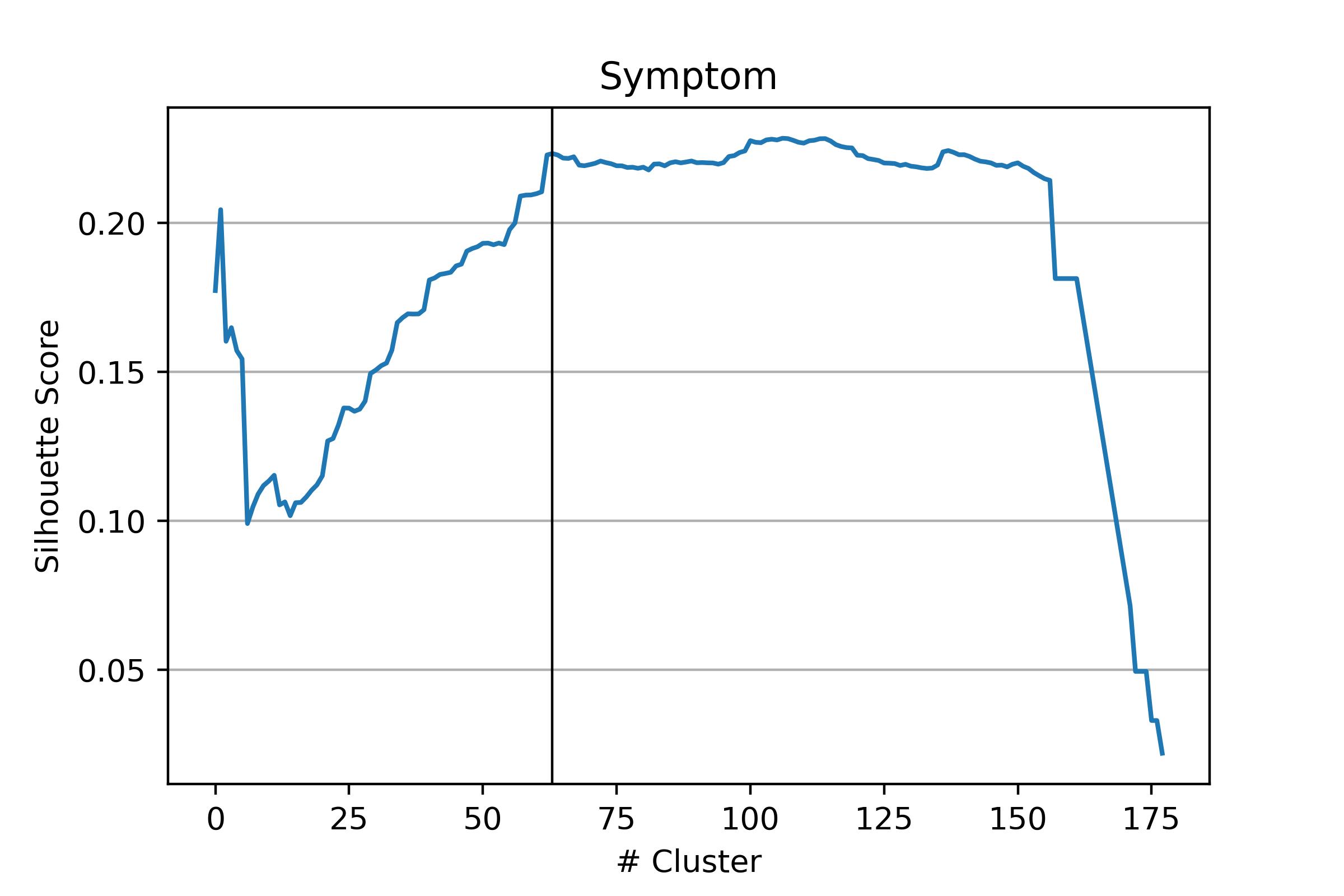}
         \caption{}
     \end{subfigure}
     \begin{subfigure}[b]{0.49\linewidth}
         \centering
         \includegraphics[width=\linewidth]{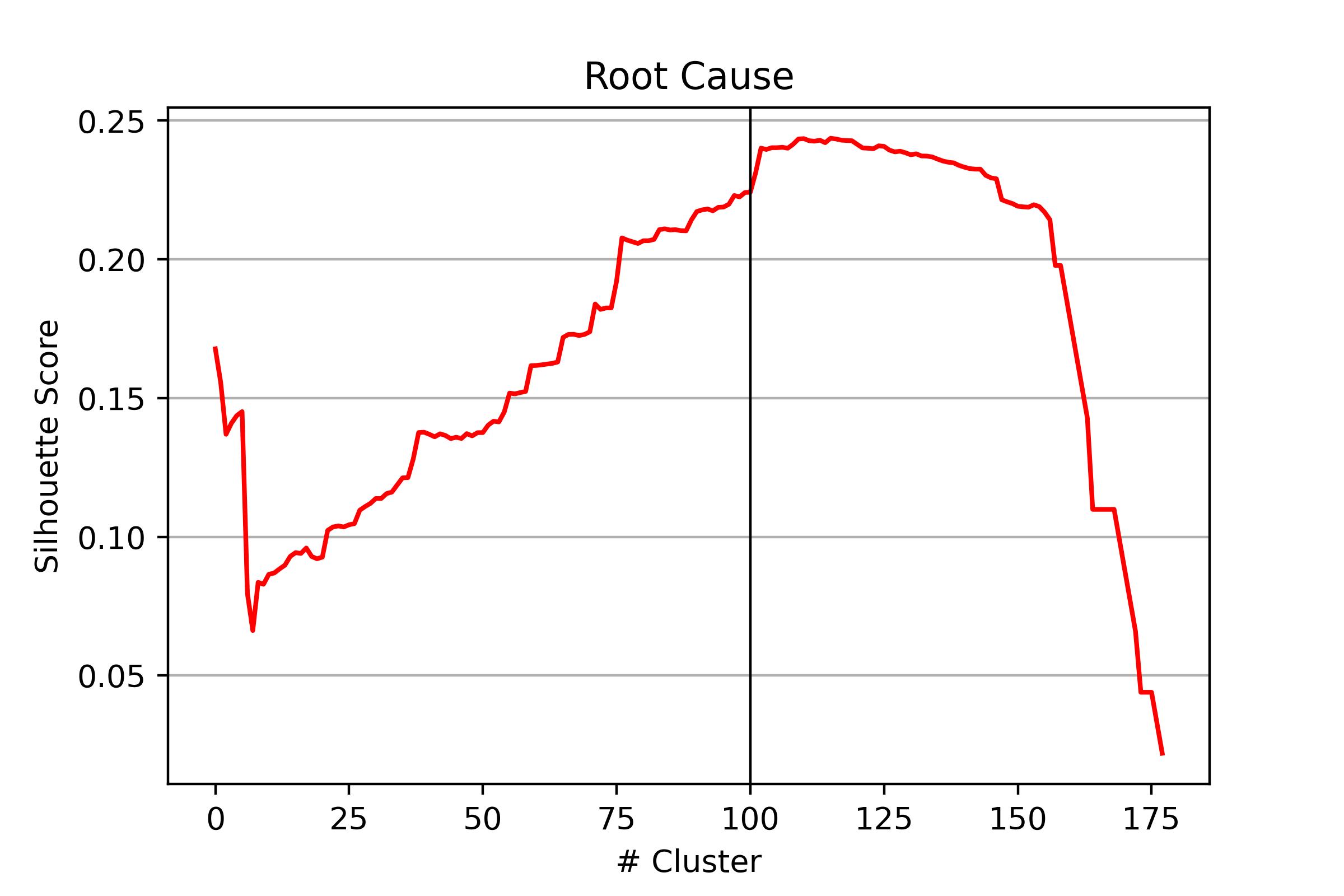}
         \caption{}
     \end{subfigure}
     \begin{subfigure}[b]{0.49\linewidth}
         \centering
         \includegraphics[width=\linewidth]{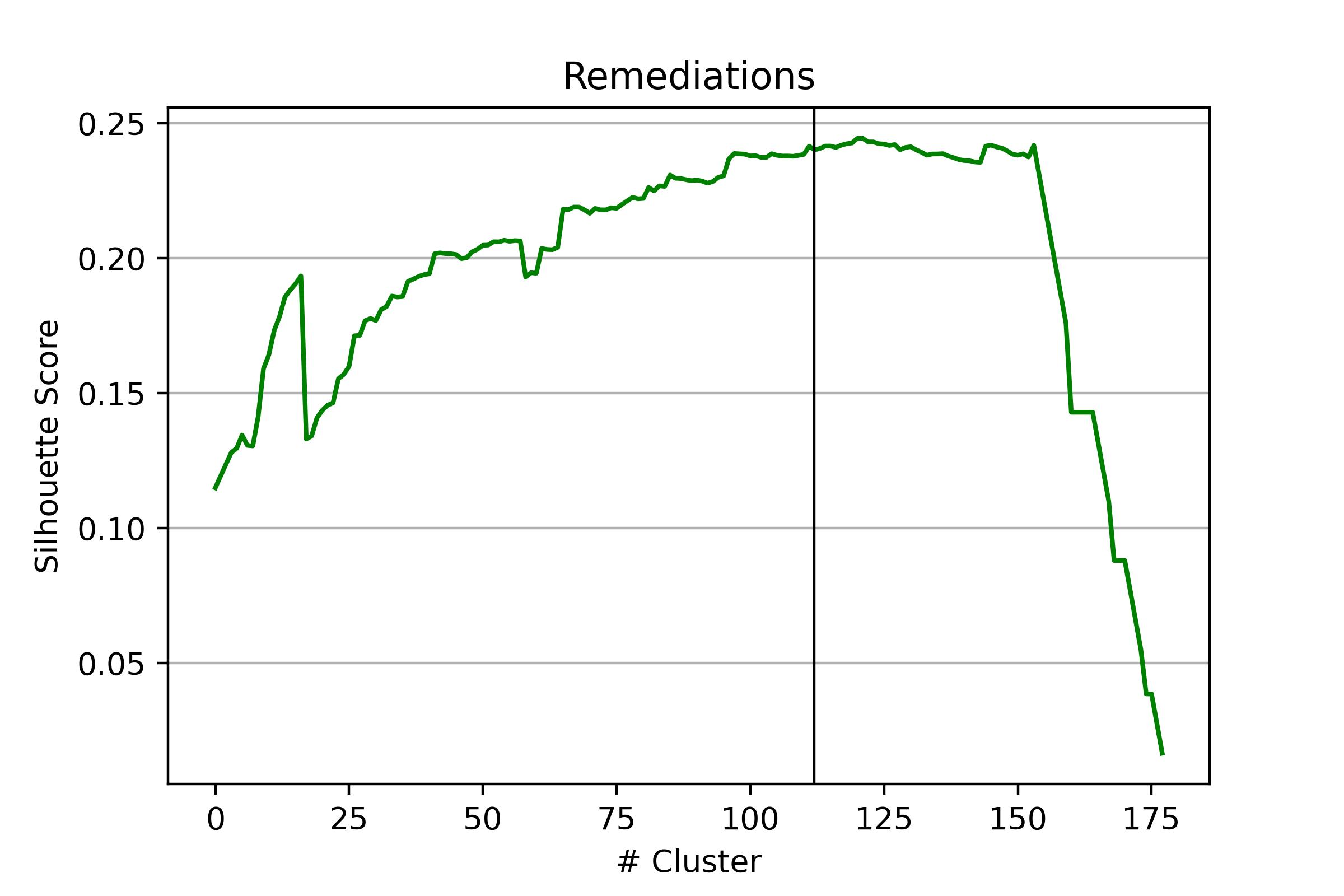}
         \caption{}
     \end{subfigure}
     \begin{subfigure}[b]{0.49\linewidth}
         \centering
         \includegraphics[width=0.7\linewidth]{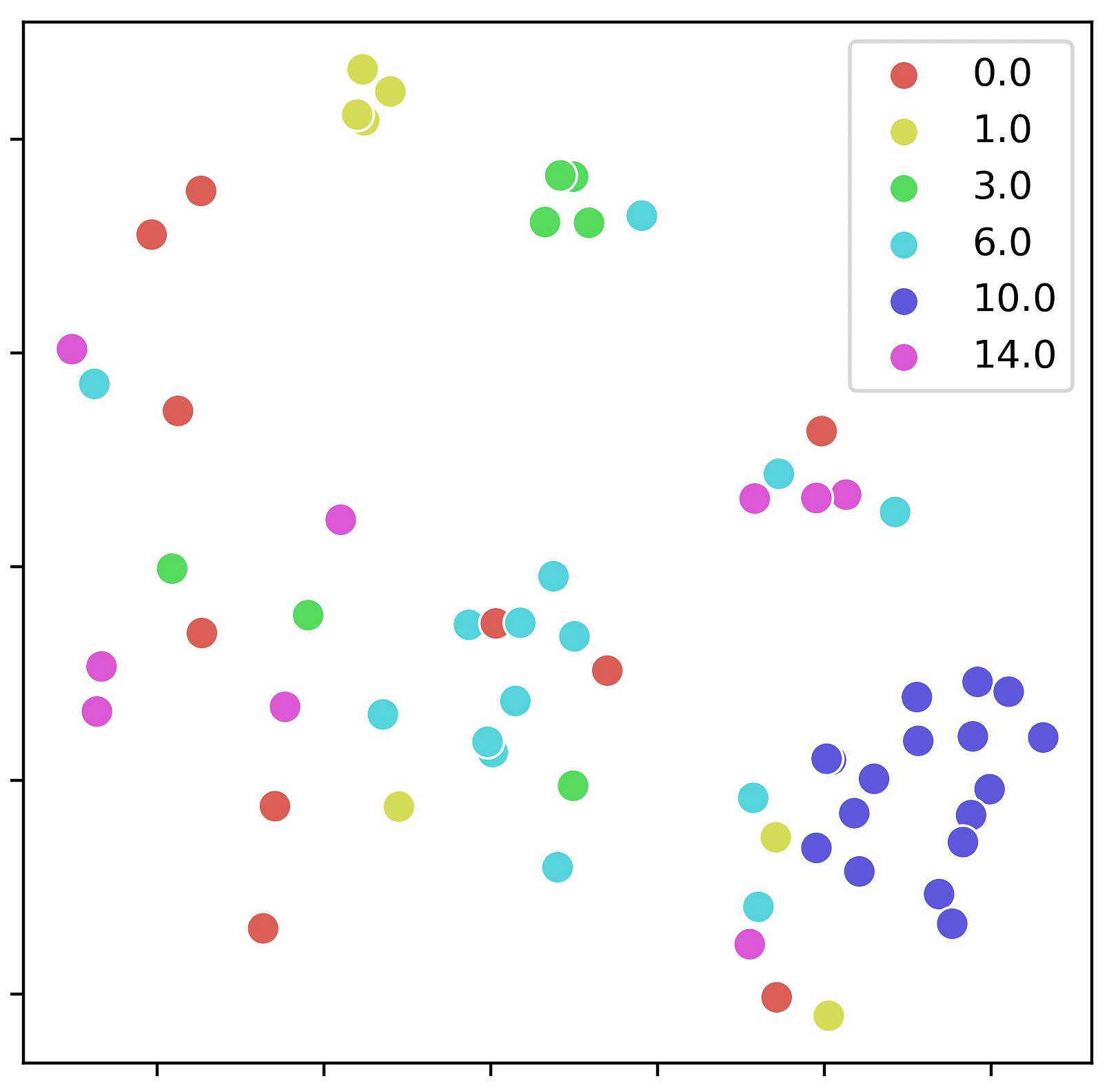}
         \caption{}
     \end{subfigure}
     \caption{\textit{Figure (a), (b) and (c) plots silhouette scores against number of clusters while clustering symptom nodes, root cause nodes and remediation nodes respectively. The vertical line represents the optimal number of clusters based on our condition (Section \S\ref{sec:kg-construction}). Figure (d) is the t-SNE visualization of the symptom embeddings for top-6 populated clusters.}}
    \label{fig:cluster_info}
\end{figure}

An important step in the graph construction pipeline for \ourmethod{} is the clustering of similar outages (Section \S\ref{sec:kg-construction}), which helps the inference method in suggesting related outages for a given alert or a set of alerts. We mentioned that the choice for the optimal number of clusters for each of the node category (symptom nodes, root cause nodes and remediation nodes) is based on the Silhouette Score computed after agglomerative clustering. We plot the variation of the silhouette scores against the number of clusters for each node category in Figure \ref{fig:cluster_info}(a), (b) and (c). Along with it, the optimal number of clusters for each category ($\mathcal{K}_{symp}$, $\mathcal{K}_{root-cause}$ and $\mathcal{K}_{rem}$) is shown through a vertical line. The optimal number of clusters are chosen such that it is within 5\% of the maximum silhouette score obtained.

With the optimal number of clusters fixed, we merge them and establish a new set of clusters such that two outages are in the same new cluster $\mathcal{K}$ if either their symptoms, root causes or remediation steps were in their same respective optimal clusters. As a result, we create 53 new set of optimal clusters that groups all the outages. In Figure \ref{fig:cluster_info}(d), we illustrate the t-SNE visualization of the symptom embeddings for the outages grouped into the top-6 largest clusters. These clusters constitute $\sim38\%$ of all the outages. We observe from the diagram that the clusters with higher number of outages (cluster 10, cluster 6, etc.) form close groups, illustrating that a new incident with a similar symptom can be associated to other past incidents.


\subsubsection{Inference Method} \label{sec:inference_method}
In this section, we draw forth a comparison between the different inference methods described in Section \S\ref{sec:inference} that utilizes the different components of the CK graph. We show how using the causal and the knowledge graph together performs better than using them individually while predicting the root cause and the remediation steps. The comparisons are reported using average results over 50 randomly selected outages (described in Section \S\ref{sec:eval_method}). The comparison will provide us with a quantitative explanation for the most suitable inference method.

\begin{table}[h]
\centering
\begin{tabular}{ccccc}
\toprule
 & \textbf{Metric} & \textbf{Path} & \textbf{Sim} & \textbf{Clust}  \\
\midrule
\textbf{Root Cause} & \textbf{Rouge-1} & 0.202 & 0.211 & 0.242  \\
\cmidrule{2-5}
& \textbf{Rouge-L} & 0.188 & 0.194 & 0.227  \\
\midrule
\textbf{Remediation} & \textbf{Rouge-1 }& 0.158 & 0.177 & 0.219  \\
\cmidrule{2-5}
& \textbf{Rouge-L} & 0.136 & 0.157 & 0.205  \\
\bottomrule
\end{tabular}
\vspace{0.3em}
\caption{\textit{Comparing the performance of the various inference methods through Rouge scores described in the paper}}
\label{tab:inference-comp}
\vspace{-3mm}
\end{table}

We observe from Table \ref{tab:inference-comp} that the clustering based inference strategy (\texttt{Clust}), as described in Section \S\ref{sec:inference_clust}, outperforms both path (\texttt{Path}) and similarity (\texttt{Sim}) based inference methods. Even though \texttt{Path} uses the causal graph to find the ancestral alerts which are directly linked to the symptom nodes, it requires knowledge graph components to predict the root cause and the remediation step to be evaluated equally. However, \texttt{Sim} uses only the knowledge graph to predict the root causes and the remediation steps. Thus, the causal graph cannot be evaluated individually, and hence we do not evaluate any baselines that uses only alerts to predict the root cause.

The evaluation using both Rough-1 and Rouge-L scores demonstrates that the \texttt{Clust} method consistently outperforms the alternative techniques. Clustering based inference method  utilizes both the sources of data to recommend potential root causes and remediation techniques. The broader scope is enabled by the utilization of the history of the outages including their diagnosis, the alerts that were triggered during an outage, and the predictive insights of each individual alert in relation to an outage scenario.
We model the indicative power of an alert through the outage cluster predictor model, which we shall further elucidate in the next section.

\subsection{Baseline Comparison}

In this section, we compare the cluster based inference method that we derived to be the best performing inference method against state-of-the-art baselines (as selected in Section \S\ref{sec:baseline}). In Table \ref{tab:baseline-comp}, we draw forth this comparison and report the results. Similar to Section \S\ref{sec:inference_method}, we report the average results over the same 50 random outages that were chosen in Section \S\ref{sec:inference_method} for the baselines.

\begin{table}[h]
\centering
\begin{tabular}{cccccc}
\toprule
 & \textbf{Metric} & \textbf{IS} & \textbf{GCN} & \textbf{Clust} & \textbf{\% Gain} \\
\midrule
\textbf{Root Cause} & \textbf{Rouge-1} & 0.207 & 0.176 & 0.242 & 27.2\% \\
\cmidrule{2-6}
 & \textbf{Rouge-L} & 0.197 & 0.165 & 0.227 & 26.4\% \\
\midrule
\textbf{Remediation} & \textbf{Rouge-1} & 0.157 & 0.162 & 0.219 & 37.3\% \\
\cmidrule{2-6}
& \textbf{Rouge-L} & 0.143 & 0.147 & 0.205 & 41.4\% \\
\bottomrule
\end{tabular}
\vspace{0.3em}
\caption{\textit{Comparison of Cluster based inference to the baselines. \% Gain indicates the average improvement over the two baselines.}}
\label{tab:baseline-comp}
\vspace{-3mm}
\end{table}

The table shows that the cluster-based inference method outperforms the Incident Search and GCN baselines. Even though these baseline methods utilized the symptom description as input, a level of detail typically available only after post-mortem reports, they still do not achieve results superior to those of the \textit{Clust} approach. 

\textit{Clust} on average exhibits 16\% higher performance in terms of Rouge-1 scores over Incident Search for root cause recommendation, while around 38\% higher for recommending remediations. Improvements in Rouge-L scores are similar as well. Meanwhile, GCN's performance is notably lower to that of Incident Search in both root cause identification and recommendation tasks. Overall, \textit{Clust} performs $\sim$27\% higher than the baselines on average in root cause recommendation and 39\% in remediation steps recommendation. 

Both the baselines use knowledge graph only to predict the root causes and the remediation steps. Incident Search mainly uses text similarity to find similar symptoms, similar to \texttt{Sim}. In addition, we use the contextualized BERT embeddings to represent the nodes of the knowledge graph, while incident search uses RoBERTa and GCN computes its own embeddings with GLoVe initialization. None of these embeddings provide additional information. An interesting observation is that the simple approach of comparing text similarities works better compared to the more complex graph-based method (GCN). This might be because our data covers a wide range of different types of outages, resulting in individual connected components for each outage report present in the data. This diversity makes it hard for graph-based methods to extract layout specific features and enrich the embedding computations.

\subsection{Outage Cluster Predictor Performance}

\begin{figure}[h]
    \centering
    \includegraphics[width=0.8\linewidth]{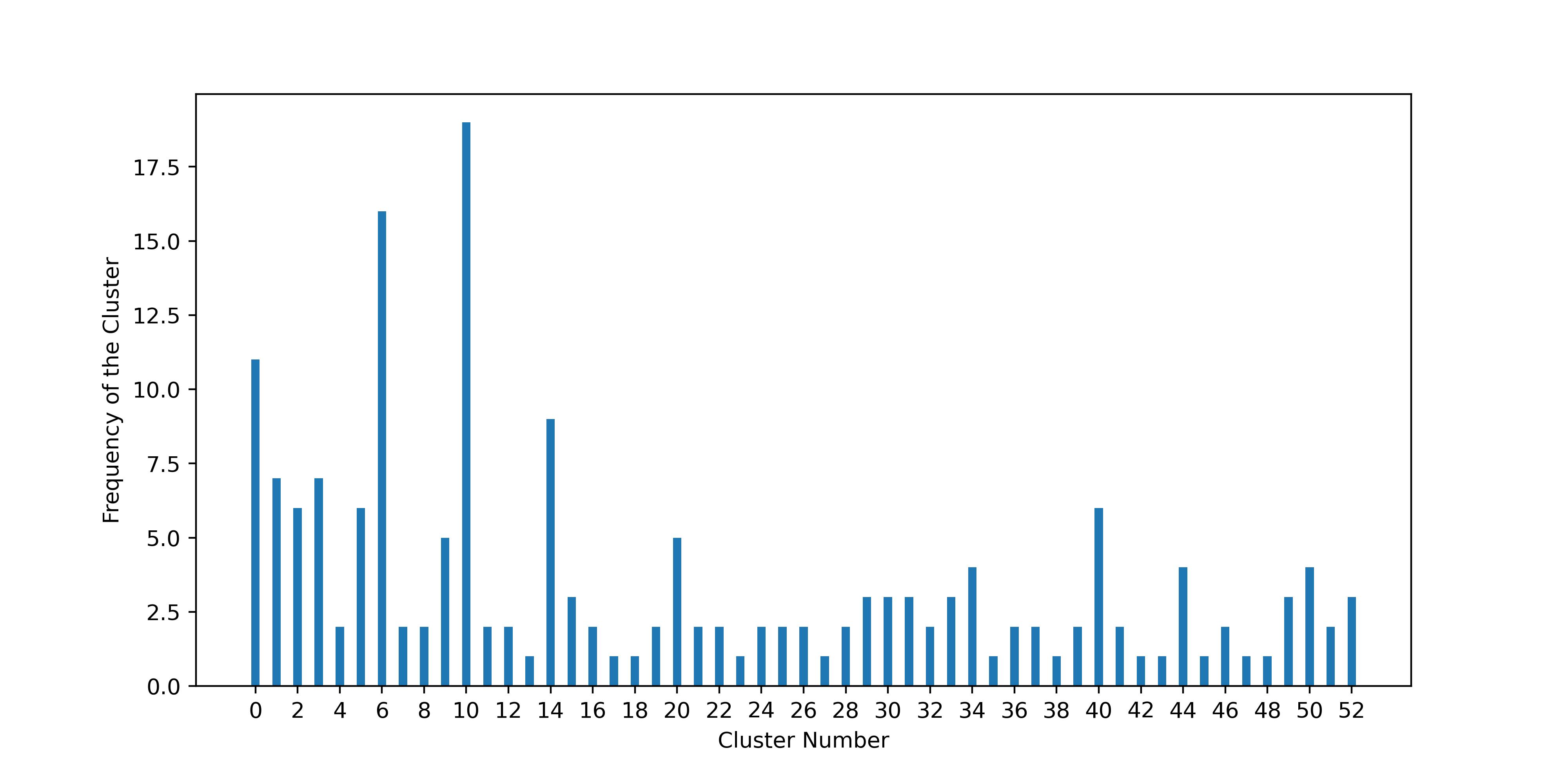}
    \caption{\textit{The figure shows the number of outages in each cluster}}
    \label{fig:cluster_freq}
    \vspace{-3mm}
\end{figure}

To compare the performance of the Outage Cluster Predictor, we plot top-K precision of the model predictions against varying K. We consider a prediction to be correct if the actual cluster is within the top-K predictions. The dataset containing all the 182 outages was split into a 70\%-30\% train-test set for training the outage cluster predictor model. The total number of clusters in the entire dataset is 53, while the train set had only 43 unique clusters. Stratified split was not possible, since few clusters represented only 1 outage (see Fig. \ref{fig:cluster_freq}).

\begin{figure}[h]
    \centering
    \includegraphics[width=0.75\linewidth]{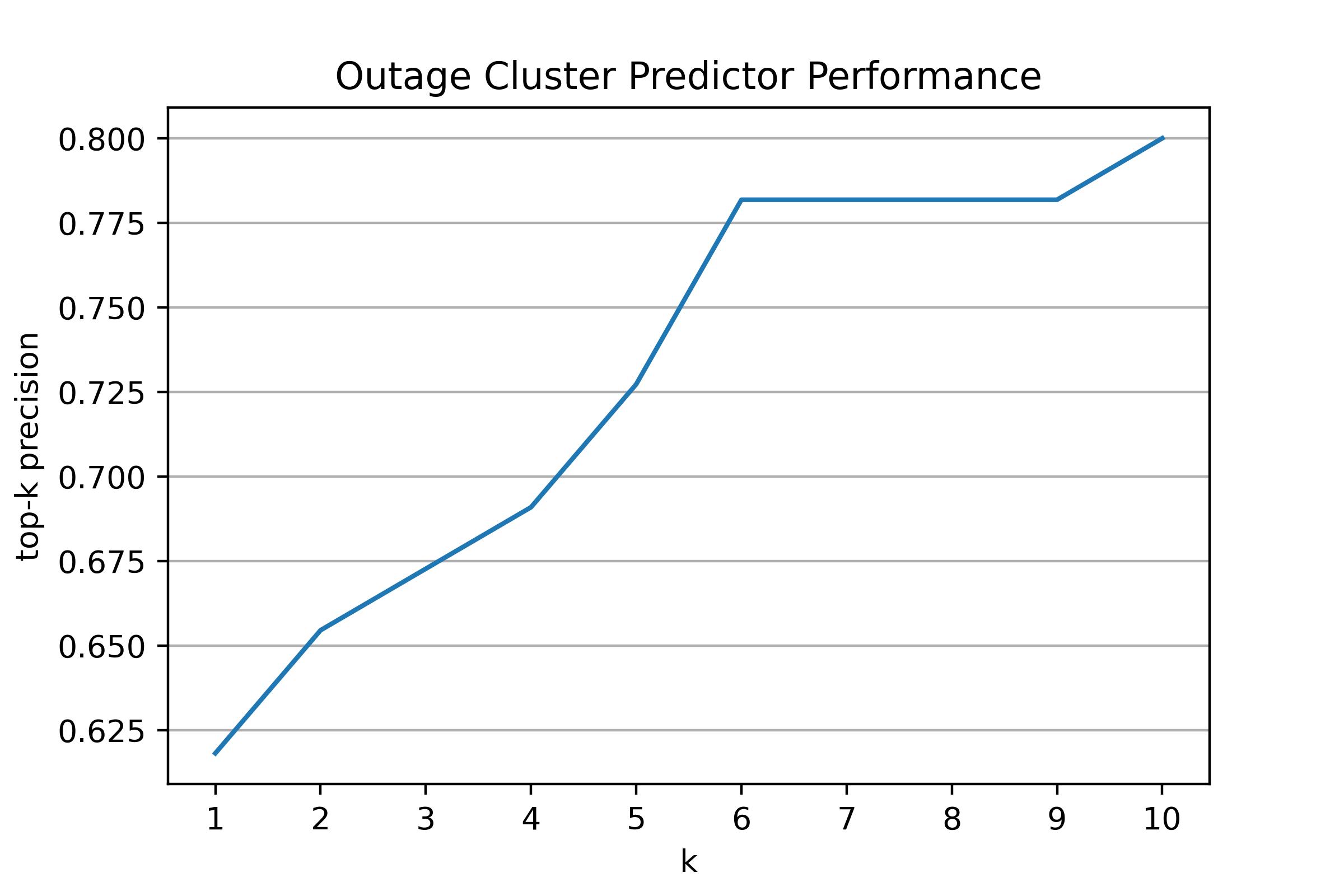}
    \caption{\textit{Test Set prediction performance of the outage cluster predictor as reported by top-K precision against varying K.}}
    \label{fig:outage_cluster_predictor}
    \vspace{-3mm}
\end{figure}

We observe in Figure \ref{fig:outage_cluster_predictor} that the top-K precision for the Outage Cluster Predictor model is 72.7\% with K=5 and over 78\% with k=6. Even with K=1, that is when we only use the top-1 prediction, the accuracy is $\sim$62\%, which shows a significant performance where the total number of available clusters are 53. 

Figure \ref{fig:cluster_freq} shows the number of outages that belong to each cluster. We see that the distribution is highly skewed with ~25\% of the outages belonging to only 3 clusters, and ~50\% of the outages belongs to only 11 clusters. Given such a skewness in the data, a top-1 accuracy of ~62\% and a top-5 accuracy of ~73\% suggests that the outage cluster predictor model is highly powerful and capable of predicting the correct cluster given the alerts that were fired for an outage.

\subsection{Illustration on Production Outages}

A quantitative evaluation captures the sentence similarity between the ground truth root causes/remediation and the predicted root cause/remediation. However, in this section, we present a formal evaluation of \ourmethod{} though manual validation of a few illustrative examples which demonstrated high Rouge scores in Section \S\ref{sec:baseline}\footnote{Showcasing qualitative examples where a high Rouge score corresponds to a strong alignment between predictions and actual outcomes}.
We present three outages that were flagged by the SREs and compare the root cause detected by the domain experts with the predicted output.

\subsubsection{Outage example 1}
This incident occurred in the email template microservice due to a deployment issue in a connected service and lasted for about 4 hours. It was caused by the service being deployed without proper configuration validation, which resulted in a fault. Running \ourmethod{} on the alerts fired during the time of the outage pointed to a similar symptom that occurred a year ago on the same email template service. It was ranked second on the list of possible prior outages. The past incident was the result of a migration of the internal deployment of the email service from one platform to another, which resulted in a change in configuration. Thus, even though there was no past outage with the same root cause among the outage reports, \ourmethod{} was able to find a fault that happened on the same email service due to a deployment issue.

\subsubsection{Outage example 2}
In this example, users of the SaaS enterprise reported an outage due to the unavailability of services that lasted approximately 1.5 hours. As per the reports, the investigations pointed to a high load in the database connectivity for the services, with the root cause identified as an inefficiency in MySQL query plan which resulted in a snowball effect. The impact was resolved by a rolling restart of the application servers as well as the deactivation of certain accounts that were the cause of the long-running database query. Executing \ourmethod{} with the alerts fired during this outage, it was able to pinpoint a similar outage 9 months before, when consumers were unable to use the same service. The root cause of the past outage was due to a resource contention at a database tier, resulting in a database connection issue. \ourmethod{} ranked the outage in top-3 among past outages. \ourmethod{} was able to find a similar symptom at the same service that was caused by a database issue.

\subsubsection{Outage example 3}
In another case, an outage was reported because a certain service was unreachable due to a deployment in the moonbeam pipeline resulting in a version mismatch. After an hour, the impact was addressed with a deployment rollback to match the versions. According to \ourmethod{}, it detected a similar outage that occurred two years ago when the same service was unreachable to a segment of users. The root cause for the past incident was a version mismatch between the service and the other components with which it communicated. To discover the version mismatch issues, a new alert was set up as a remediation strategy. Based on the similarities of the alerts fired and the previous symptoms, \ourmethod{} was able to identify this previous outage, ranking it in top-1 among similar outages.

%% file: tex/10.Conclusion.tex
\section{Conclusion}
In this work, we introduce \ourmethod{}, a novel service for identifying root causes and recommending mitigation steps during outages. By analysing semi-structured natural language text from past outage reports, as well as real-time alerts, \ourmethod{} leverages a merged graph that combines causal and knowledge graph components. This enables the capture of causal relationships among alerts and information about symptoms, root causes, and remediation techniques from past incidents. Grouping similar outages into clusters based on shared characteristics further refines our approach.
During the inference phase, we utilize the graph to discern probable root causes and potential remediation methods by drawing on insights from past outage incidents. Our  cluster-based inference method employs only outage-specific alerts to uncover similar incidents and provide root cause solutions. Through qualitative analysis and quantitative evaluation on  real outage examples, using two years of cloud service outage data, we showcase \ourmethod{}'s effectiveness. We achieve more than a 26\% enhancement in root cause prediction and over 37\% improvement in remediation step prediction compared to baseline methods.


\textbf{Future Works:}
We plan to extend our approach by conducting a broader comparative analysis by modifying state-of-the-art techniques for comparable evaluation and incorporating additional metrics for comprehensive assessment. We also propose exploring hierarchical models to predict root cause at multiple hierarchies. To make the system further intuitive to be used, investigating the utilization of LLMs for question answering within the merged CK graph holds promise. 

\clearpage